\documentclass[aps,prd,          groupedaddress,amsmath,amssymb]{revtex4}

\usepackage{epsfig}
\usepackage{dcolumn}
\usepackage{bm}

\begin{document}

\def\as{\alpha_s}
\def\bbar{\bar{b}}
\def\d{{\rm d}}
\def\eps{\varepsilon}
\def\nbar{\bar{N}}
\def\nn{\nonumber}
\def\lp{\left. }
\def\rp{\right. }
\def\lr{\left( }
\def\rr{\right) }
\def\le{\left[ }
\def\re{\right] }
\def\lg{\left\{ }
\def\rg{\right\} }
\def\lb{\left| }
\def\rb{\right| }
\def\beq{\begin{equation}}
\def\eeq{\end{equation}}
\def\bea{\begin{eqnarray}}
\def\eea{\end{eqnarray}}

\preprint{FREIBURG PHENO-07-05}
\preprint{KA-TP-18-2007}
\preprint{LPSC 07-122}
\preprint{SFB/CPP-07-59}
\title{~\\~\\~\\
 Precision predictions for $Z'$-production at the CERN LHC: \\
 QCD matrix elements, parton showers, and joint resummation}
\author{Benjamin Fuks$^{a,b}$}
\author{Michael Klasen$^b$}
\email[]{klasen@lpsc.in2p3.fr}
\author{Fabienne Ledroit$^b$}
\author{Qiang Li$^{b,c}$}
\author{Julien Morel$^b$}
\affiliation{$^a$ Physikalisches Institut, Albert-Ludwigs-Universit\"at
 Freiburg, Hermann-Herder-Stra\ss{}e 3, D-79106 Freiburg i.Br.,
 Germany \\
 $^b$ Laboratoire de Physique Subatomique et de Cosmologie,
 Universit\'e Joseph Fourier/CNRS-IN2P3/INPG,
 53 Avenue des Martyrs, F-38026 Grenoble, France \\
 $^c$ Institut f\"ur Theoretische Physik, Universit\"at
 Karlsruhe, Postfach 6980, D-76128 Karlsruhe, Germany}
\date{\today}
\begin{abstract}
We improve the theoretical predictions for the production of extra neutral
gauge bosons at hadron colliders by implementing the $Z'$ bosons in the MC@NLO
generator and by computing their differential and total cross sections in
joint $p_T$ and threshold resummation. The two improved predictions are
found to be in excellent agreement with each other for mass spectra, $p_T$
spectra, and total cross sections, while the PYTHIA parton and ``power''
shower predictions usually employed for experimental analyses show
significant shortcomings both in normalization and shape. The theoretical
uncertainties from scale and parton density variations and non-perturbative
effects are found to be 9\%, 8\%, and less than 5\%, respectively, and thus
under good
control. The implementation of our improved predictions in terms of the
new MC@NLO generator or resummed $K$ factors in the analysis chains of the
Tevatron and LHC experiments should be straightforward and lead to more
precise determinations or limits of the $Z'$ boson masses and/or couplings.
\end{abstract}
\maketitle


\vspace*{-108mm}
\noindent FREIBURG PHENO-07-05\\
\noindent KA-TP-18-2007\\
\noindent LPSC 07-122\\
\noindent SFB/CPP-07-59\\
\vspace*{83mm}

\section{Introduction}
\label{sec:1}

Despite its impressive phenomenological success, the Standard Model (SM) of
particle physics is widely believed to suffer from a variety of conceptual
deficiencies. In particular, it provides no fundamental motivation why the
strong and electroweak interactions should be described by three different
gauge groups, i.e.\ $SU(3)$, $SU(2)$, and $U(1)$. Grand Unified Theories
(GUTs) allow for a unification of these groups within a simple Lie group
such as $SU(5)$, $SO(10)$, or $E_6$. Depending on the rank of the unifying
group, one or several extra neutral gauge ($Z'$) bosons appear when the
unification group is broken to the SM at higher scales, exhibiting the
existence of additional $U(1)$ symmetries \cite{Leike:1998wr}. Anomaly
cancellations and gauge invariance of quark and lepton Yukawa couplings
impose a number of restrictions on these additional symmetries. Viable
families of models, that are consistent with constraints coming from the
CERN LEP collider, include those based on the $B-L$ and $SO(10)$ symmetries
and, provided that fermion mass generation is not restricted to the SM Higgs
mechanism and new charged fermions are allowed, also those inspired by $E_6$
\cite{Carena:2004xs}.

If the extra $Z'$-bosons couple to quarks and leptons with approximately SM
strength and if their mass is not too large, they will be produced at
current and future hadron colliders and can be easily detected through their
leptonic decay channels. The search for these particles occupies therefore
an important place in the experimental programs of the Fermilab Tevatron and
the CERN LHC. For example, the CDF collaboration has searched the Tevatron
Run~II data for $Z'$-bosons in the $e^+e^-$ decay channel, using the
di-electron invariant mass and angular distributions and setting lower mass
limits of 650 to 900 GeV for a large variety of models
\cite{Abulencia:2006iv}. Similar constraints come from electroweak precision
fits and di-fermion production at LEP2 \cite{lepewwg}. Within the ATLAS
collaboration, the discovery
reach in $Z'\to e^+e^-$ decays has recently been analyzed \cite{ledroit} for
the four classes of models defined in Ref.\ \cite{Carena:2004xs}. The CMS
collaboration has claimed a discovery reach of masses between 3.4 and 4.3
TeV for the $Z'\to\mu^+\mu^-$ decay channel and an integrated luminosity of
100 fb$^{-1}$ \cite{cousins}.

The currently available simulations for the LHC experiments rely completely
on the PYTHIA Monte Carlo (MC) generator \cite{Sjostrand:2006za}, which is
based on leading order (LO) QCD matrix elements, parton showers, and the
Lund string hadronization model. It includes the full interference structure
of new $Z'$-bosons with Drell-Yan photon and SM $Z$-boson exchange, and the
above-mentioned phenomenologically viable models can easily be implemented
\cite{Ciobanu:2005pv}. The description of the transverse-momentum ($p_T$)
spectrum of the produced vector-boson can be improved by matching the parton
shower to the hard emission of an extra parton \cite{Miu:1998ju}, but the
overall normalization of the theoretical cross section remains subject to
large higher-order corrections and scale uncertainties. The CDF
collaboration has therefore chosen to renormalize the generated Monte Carlo
events with a correction ($K$) factor in each invariant mass bin to the
next-to-next-to-leading order (NNLO) cross section \cite{Hamberg:1990np}.
However, this procedure does not lead to a correct description of the
transverse-momentum spectrum. Note also that in principle the LO cross
section no longer factorizes at NNLO, i.e.\ the $K$-factors for Drell-Yan
and $Z'$ production need no longer to be equal \cite{Carena:2004xs}.

Here, we report on the implementation of extra $Z'$-bosons in the
next-to-leading order (NLO) Monte Carlo generator MC@NLO
\cite{Frixione:2002ik}, allowing to match the complete NLO matrix elements
with the parton shower and cluster hadronization model of the Monte Carlo
generator HERWIG \cite{Corcella:2000bw}. Since the LO cross section still
factorizes completely at NLO, this requires the implementation of the
$Z'$-boson mass, decay width, propagator, and couplings in MC@NLO, together
with the full interference with Drell-Yan photon and SM $Z$-boson exchanges.
As was the case for PYTHIA, the emission of one additional hard parton has
previously been matched to the HERWIG parton shower, albeit for SM
vector-boson production only \cite{Corcella:1999gs}. As an alternative, the
CKKW formalism \cite{Catani:2001cc} for the matching of hard real emissions
has been implemented in both PYTHIA and HERWIG for SM vector-boson
production with the result that the three different matching formalisms were
found to vary systematically in a significant way and to depend in addition
strongly on the matching scale \cite{Mrenna:2003if}. Similar results were
obtained more recently in \cite{Alwall:2007fs} for $W$-boson plus jet
production, where in addition various implementations of the MLM
\cite{Caravaglios:1998yr} prescription for matching multiparton final states
to parton showers have been compared to the methods discussed above. Here,
we perform a
similar study of systematic uncertainties by comparing the invariant-mass
and transverse-momentum distributions as predicted with our implementation
of $Z'$-bosons in MC@NLO with the PYTHIA parton-shower and matrix-element
corrections. We also confront these MC predictions with a new computation of
$Z'$-boson production in the framework of joint resummation
\cite{Bozzi:2007te}. In addition, we compare the dependence of the
various predictions on the unphysical renormalization and factorization
scales as well as on the employed parton densities. The impact of
hadronization corrections, dominant electroweak corrections, and
non-perturbative effects are also studied.

The remainder of this work is organized as follows: in Sec.\ \ref{sec:2}
we first describe the PYTHIA framework for $Z'$-boson production and the
associated matching of parton showers with the emission of an additional
hard parton. We then discuss our implementation of $Z'$-boson production in
the MC@NLO generator and recall its matching procedure of NLO matrix
elements with the HERWIG mechanism of parton showers. We also present
briefly our calculation of $Z'$-boson production using joint resummation.
In Sec.\ \ref{sec:3}, we define our choice of electroweak SM parameters and
define the parameters of our exemplary $Z'$ model. We also discuss the
various corrections that we apply to the production cross section, i.e.\
dominant electroweak corrections, next-to-leading order QCD matrix elements,
parton showers, and hadronization. We then compare the numerical impact of
these corrections and study the remaining theoretical uncertainties, coming
from the choice of renormalization and factorization scales, the
parameterization of parton densities, and non-perturbative effects. Our
conclusions are given in Sec.\ \ref{sec:4}.

\section{Theoretical setup}
\label{sec:2}

In 1984, Green and Schwarz showed that ten-dimensional string theories with
$E_8\times E_8$ or $SO(32)$ gauge symmetry are anomaly-free and thus
potentially finite \cite{Green:1984sg}. Among these two gauge groups, only
the former contains chiral fermions as they exist in the SM. After
compactification, it leads to the $E_6$ symmetry as an effective GUT group,
that can be broken further to \cite{Hewett:1988xc}
\beq
 E_6\,\to\,SO(10) \times U(1)_\psi
    \,\to\,SU(5)  \times U(1)_\chi \times U(1)_\psi.
 \label{eq:1}
\eeq
While the $Z'$-bosons corresponding to the additional $U(1)_\psi$ and
$U(1)_\chi$ symmetries can in general mix with each other,
\beq
 Z' (\theta)\,=\,Z_\psi\cos\theta+Z_\chi\sin\theta ~~~{\rm and}~~~
 Z''(\theta)\,=\,Z_\psi\sin\theta-Z_\chi\cos\theta,
\eeq
we consider in this work only a TeV-scale $Z_\chi$-boson as an exemplary
case, i.e.\ $\theta=90^\circ$ in the convention of Ref.\
\cite{Rosner:1986cv}, and assume the $Z''\equiv Z_\psi$ to acquire its mass
at considerably higher scales, as it is naturally the case in the hierarchy
of symmetry breaking of Eq.\ (\ref{eq:1}). We will furthermore assume that
$SO(10)$ breaks down to $SU(5)\times U(1)_\chi$ at the same scale where
$SU(5)$ breaks down to the SM group $SU(3)_C\times SU(2)_L\times U(1)_Y$
with gauge couplings $g_s$, $g$, and $g'$. The $U(1)_\chi$ coupling
\beq
 g_\chi\,=\,
 \sqrt{5\over3}\,g'\,=\,
 \sqrt{5\over3}\,g\,\tan\theta_W\,=\,
 \sqrt{5\over3}\,{e\over\sqrt{1-\sin^2\theta_W}}
 \label{eq:3}
\eeq
is then directly related to the coupling $g'$ of the weak hypercharge
$U(1)_Y$ by the usual group-theoretical factor $\sqrt{5/3}$. Note that as
the $SO(10)$-breaking scale increases from the $SU(5)$-breaking scale to the
$E_6$ or Planck scale, $g_\chi$ starts to deviate from $\sqrt{5/3}\,g'$ to
roughly $\sqrt{2/3}$ times this value \cite{Robinett:1981yz}. Using Eq.\
(\ref{eq:3}), we can thus express $g_\chi$ in terms of a group-theoretical
factor, the $SU(2)_L$ gauge coupling $g$ or alternatively the
electromagnetic fine structure constant $\alpha=e^2/(4\pi)$, and the squared
sine of the angle $\theta_W$, describing the mixing of the neutral $W^0$-
and $B$-bosons to the massless photon and the SM $Z$-boson of mass $m_Z$
after the breaking of $SU(2)_L\times U(1)_Y$ to $U(1)_{\rm em.}$. While the
photon is protected from further mixing by the exact electromagnetic
symmetry, the SM $Z$-boson can in general mix with the additional
$Z'$-boson with an angle $\phi$. Its squared tangent
\beq
 \tan^2\phi\,=\,{m_Z^2-m_1^2\over m_2^2-m_Z^2}
\eeq
depends then on the eigenvalues of the $Z-Z'$ mass matrix $m_{1,2}$, which
in turn depend on the vacuum expectation values $v_{10}$ and $v_{\rm SM}$ of
the $SO(10)$- and SM-breaking Higgs fields. This mixing can in general
induce a coupling of the $Z'$-boson to the SM $W^\pm$-bosons, even though
they belong to different gauge groups. However, the ratio $v_{10}/v_{\rm
SM}$ is usually large and $m_1\simeq m_Z\,\ll\, m_2\simeq m_{Z'}$, so that
$Z-Z'$ mixing will be neglected in the following. The DELPHI collaboration
has constrained $\phi$ to be smaller than a few mrad ($1.7$ mrad for
$m_{Z'}=440$ GeV in the $\chi$-model) \cite{Abreu:2000ap}.

\subsection{$Z'$ production in PYTHIA}
\label{sec:2.1}

In PYTHIA 6.403 \cite{Sjostrand:2006za}, the production of extra neutral
gauge bosons in hadron collisions has been implemented including the full
interference structure with SM photon and $Z$-boson exchanges. The
Lagrangian describing the interaction of the extra neutral gauge boson $Z'$
with fermions $f$
\beq
 {\cal L}\,=\,{g\over4\cos\theta_W}\bar{f}\gamma^\mu(v_f-a_f\gamma_5)f
 Z'_\mu \label{eq:5}
\eeq
has been expressed in terms of generalized vector ($v_f$) and axial-vector
couplings ($a_f$), which depend in general on the unifying gauge group and
the Higgs representations employed to break this group down to the SM gauge
group. For the additional $U(1)_\chi$ symmetry, that we use as our standard
example, these couplings are given in Tab.\ \ref{tab:1} for the down- and
\begin{table}
 \begin{center}
 \caption{Vector and axial-vector couplings of down- and up-type quarks as
 well as charged leptons and neutrinos in the $E_6$-inspired $Z_\chi$
 model in the convention of PYTHIA. $s_W$ is the sine of the electroweak
 mixing angle $\theta_W$.\\
 \label{tab:1}}
 \begin{tabular}{|cc|cc|cc|cc|}
 \hline
 $v_d$ &  $a_d$ &  $v_u$ &  $a_u$ &  $v_l$ &  $a_l$ &  $v_\nu$ &  $a_\nu$ \\
 \hline
 $2\sqrt{6} s_W/3$ & $-\sqrt{6} s_W/3$ & $0$ & $\sqrt{6} s_W/3$ & $-2\sqrt{6} s_W/3$ & $-\sqrt{6} s_W/3$ & $-\sqrt{6} s_W/2$ & $-\sqrt{6} s_W/2$ \\
 \hline
 \end{tabular}
 \end{center}
\end{table}
up-type quarks as well as for the charged leptons and neutrinos. We will
be interested in the Drell-Yan like production of electron-positron pairs
\beq
 q\,\bar{q}\,\to\,(\gamma,Z,Z')\,\to\, e^-\,e^+,
 \label{eq:6}
\eeq
i.e.\ the relevant couplings are primarily those of the five light quark
flavours $q=u,d,s,c,b$ (with masses $m_q$ much smaller than the partonic
centre-of-mass energy $\sqrt{s}$) and positrons/electrons $e^\pm$ and to
a lesser extent those of the other fermions, which contribute to the total
decay width appearing in the $Z'$ propagator.

\subsection{Matching of parton showers with LO matrix elements in PYTHIA}

The only part of the Drell-Yan like process in Eq.\ (\ref{eq:6}) that is
sensitive to QCD corrections is the quark-antiquark (and beyond the LO the
quark-gluon) initial state. It can
give rise to an initial-state parton shower, that is modeled in PYTHIA by
starting with the hard scattering partons and then successively
reconstructing the preceding branchings in a falling sequence of spacelike
virtualities $Q^2$. They range from a maximal value $Q_{\max}=m_{Z^{(\prime)
}}$, that is of the order of the hard scattering scale, to a cut-off scale
$Q_0=1$ GeV, that is close to a typical hadron mass.

The scale $Q_{\max}$ is, however, not uniquely defined. It can in
particular be increased to the hadronic centre-of-mass energy $\sqrt{S}$, so
that the parton shower (PS) populates the full phase space. However, it must
then be matched to the QCD matrix element (ME) describing the emission of
one extra hard parton
\beq
 q\,\bar{q}\,\to\,(\gamma,Z,Z')\,g
 \label{eq:8}
\eeq
at ${\cal O}(\alpha\alpha_s)$. The emission rate for the final (and normally
hardest) $q\to qg$ emission must therefore be corrected by a factor
\cite{Miu:1998ju}
\beq
 R_{q\,\bar{q}\,\to\,(\gamma,Z,Z')\,g}(s,t)\,=\,
 {(\d\sigma/\d t)_{\rm ME}\over(\d\sigma/\d t)_{\rm PS}}\,=\,
 {\sum_{i=\gamma,Z,Z'}[t^2+u^2+2m_i^2s]+{\rm interference~terms}\over
  \sum_{i=\gamma,Z,Z'}[s^2+m_i^4]+{\rm interference~terms}},
\eeq
which always lies between one-half and one. Here, $s$, $t$ and $u$ refer to
the usual Mandelstam variables of the process in Eq.\ (\ref{eq:8}), and
$m_\gamma=0$. At the same order, the crossed process
\beq
 q\,g\,\to\,(\gamma,Z,Z')\,q
\eeq
can occur, in which case the correction factor is
\beq
 R_{q\,g\,\to\,V\,q}(s,t)\,=\,
 {(\d\sigma/\d t)_{\rm ME}\over(\d\sigma/\d t)_{\rm PS}}\,=\,
 {\sum_{i=\gamma,Z,Z'}[s^2+u^2+2m_i^2t]+{\rm interference~terms}\over
  \sum_{i=\gamma,Z,Z'}[(s-m_i^2)^2+m_i^4]+{\rm interference~terms}}.
\eeq
Since this factor always lies between one and three, the $g\to q\bar{q}$
splitting must be preweighted by a factor three in order to correctly
reproduce the $s$-channel graph $q\,g\to q^*\to(\gamma,Z,Z')\,q$.
Besides the dominant QCD radiation, we will also briefly investigate the
effect of the relatively suppressed QED radiation (e.g.\ $q\to q\gamma$) as
well as the corrections induced by the hadronization of the additional
partons.

In addition to the transverse momentum $p_T$ generated by hard emission
and/or the initial-state parton shower, an average intrinsic
transverse-momentum $\langle k_T\rangle$ can be assigned to the shower
initiator in order to take into account the transverse motion of quarks
inside the original hadron. As the shower does not evolve below $Q_0=1$ GeV,
this same value is retained in PYTHIA as the default value for $\langle k_T
\rangle$. However, just like $Q_{\max}$, $Q_0\equiv\langle k_T\rangle$ is
not uniquely defined and furthermore closely related to the non-perturbative
regime of QCD. We therefore set $\langle k_T \rangle=0$ in the following.

PYTHIA allows in principle also for the participation of multiple parton
pairs in hadronic collisions. We do, however, not make use of this
possibility, as it has little numerical effect and its description in
perturbative QCD remains controversial.

\subsection{Implementation of $Z'$-bosons in MC@NLO}

In MC@NLO \cite{Frixione:2002ik}, the implementation of SM $Z$-boson
interactions with fermions $f$ is based on the Lagrangian
\cite{Aurenche:1980tp}
\beq
 {g\over\cos\theta_W}\bar{f}\gamma^\mu(a_f+b_f\gamma_5)fZ_\mu.
\eeq
For photon interactions, $a_f=e_f\sin\theta_W\cos\theta_W$, where $e_f$ is
the fractional fermion charge, and $b_f=0$. The vector coupling $a_f$ and
the axial vector coupling $b_f$ are related to those defined in PYTHIA (see
Eq.\ (\ref{eq:5})) by $a_f\to v_f/4$ and $b_f\to -a_f/4$. They are combined
to form the coefficients $A_f\,=\,a_f^2+b_f^2$ and $B_f\,=\,2a_fb_f$, that
appear in the squared matrix elements
\bea
 \overline{|{\cal M}_i|^2}(q\bar{q}~{\rm or}~qg\to\gamma,Z\to e^-e^++X)&=&
 {1\over4}\,e^4\,C_i\,\lg {e_q^2\over M^4}T_i|^{1,0}_{1,0}\rp\nonumber\\ &+&
 {1\over\sin^4\theta_W\cos^4\theta_W} {1\over(M^2-m_Z^2)^2+(\Gamma_Zm_Z)^2}
 T_i|^{A_l,B_l}_{A_q,B_q} \nonumber\\ &-&
\lp {2e_q\over M^2}
 {1\over\sin^2\theta_W\cos^2\theta_W} {M^2-m_Z^2\over(M^2-m_Z^2)^2
 +(\Gamma_Zm_Z)^2} T_i|^{a_l,b_l}_{a_q,b_q}\rg,
\eea
that have been averaged/summed over initial/final spins and colours. For the
LO Drell-Yan (DY) process in Eq.\ (\ref{eq:6}), the colour factor is $C_{\rm
DY}={N_C/N_C^2}=1/3$ and
\beq
 T_{\rm DY}|^{A_l,B_l}_{A_q,B_q}\,=\,
 8\le A_l A_q \,(t_{\rm DY}^2+u_{\rm DY}^2)
     -B_l B_q \,(t_{\rm DY}^2-u_{\rm DY}^2)\re,
\eeq
where $M^2=s_{\rm DY}$ is the invariant mass of the lepton pair and $s_{\rm
DY}$, $t_{\rm DY}$, and $u_{\rm DY}$ refer to the Mandelstam variables of
the DY-process in Eq.\ (\ref{eq:6}). The NLO QCD-corrections to the
DY-process have been implemented in MC@NLO using the same convention
\cite{Aurenche:1980tp}.

One major new and technical aspect of our work is the implementation of
$Z'$-boson interactions and exchanges in the framework described above. To
this end, we have defined the mass and width of the $Z'$-boson as well as
its couplings in the convention of PYTHIA. The squared matrix element has
also been modified,
\bea
 \overline{|{\cal M}_i|^2}(q\bar{q}~{\rm or}~qg\to&\gamma,Z,Z'&\to e^-e^++X)
 ~=~ {1\over4}\,e^4\,C_i\,\lg {e_q^2\over M^4}T_i|^{1,0}_{1,0}\rp\nonumber\\
 &+& {1\over\sin^4\theta_W\cos^4\theta_W} {1\over(M^2-m_Z^2)^2+(\Gamma_Zm_Z)^2} T_i|^{A_l,B_l}_{A_q,B_q} \nonumber\\
 &+& {1\over\sin^4\theta_W\cos^4\theta_W} {1\over(M^2-m_{Z'}^2)^2+(\Gamma_{Z'}m_{Z'})^2} T_i|^{A'_l,B'_l}_{A'_q,B'_q} \nonumber\\
 &-& {2e_q\over M^2} {1\over\sin^2\theta_W\cos^2\theta_W} {M^2-m_Z   ^2\over(M^2-m_Z   ^2)^2+(\Gamma_Z   m_Z   )^2} T_i|^{a_l,b_l}_{a_q,b_q}\nonumber\\
 &-& {2e_q\over M^2} {1\over\sin^2\theta_W\cos^2\theta_W} {M^2-m_{Z'}^2\over(M^2-m_{Z'}^2)^2+(\Gamma_{Z'}m_{Z'})^2} T_i|^{a'_l,b'_l}_{a'_q,b'_q}\nonumber\\
 &+&  2              {1\over\sin^4\theta_W\cos^4\theta_W} {(M^2-m_Z^2) (M^2-m_{Z'}^2)+\Gamma_Z m_Z \Gamma_{Z'} m_{Z'} \over [(M^2-m_Z^2)^2+(\Gamma_Zm_Z)^2] \times [(M^2-m_{Z'}^2)^2+(\Gamma_{Z'}m_{Z'})^2]} \nonumber\\
 &\times& \lp T_i|^{a_l\,a'_l+b_l\,b'_l\,,a_l\,b'_l+a'_l\,b_l}_{a_qa'_q+b_qb'_q,a_qb'_q+a'_qb_q}\rg.
\eea
It includes now the squared $Z'$-boson exchange as well as its interferences
with the photon and SM $Z$-boson exchanges.
Note that it is not sufficient to change only the $Z$-boson mass in the
existing MC@NLO implementation, but that it is also necessary to change the
$Z$-boson width as well as the couplings. As a consequence, all observables
change: the mass spectrum due to the modified width, the forward-backward
asymmetries due to the modified couplings, etc. Furthermore, even if one
changes all of these parameters, one still has $\gamma-Z'$ interference, but
no $Z-Z'$ interference. We have checked numerically that these modifications
induce large differences already in the total cross section.

\subsection{Matching of parton showers with NLO matrix elements in MC@NLO}

In HERWIG \cite{Corcella:2000bw}, parton showers are generated by a coherent
branching algorithm with parton splittings $i\to jk$, whose energy fractions
$z_j=E_j/E_i$ are distributed according to the LO DGLAP splitting functions.
Phase space is restricted to an angular ordered region, which automatically
takes infrared singularities into account, i.e.\ at each branching, the
angle between the two emitted partons is smaller than that of the previous
branching. The emission angles $\xi_{jk}=(p_j.p_k)/(E_j E_k)\simeq{1\over2}
\theta_{jk}^2$ are distributed according to Sudakov form factors, that sum
virtual corrections and unresolved real emissions and normalize the
branching distributions to give the probabilistic interpretation needed for
a MC simulation. For initial-state radiation, the parton shower follows, of
course, a backward evolution. It is terminated when $\xi_{jk}<Q_0^2/E_i^2$,
where the space-like cut-off scale $Q_0$ is set to 2.5 GeV by default.
Below this scale, a non-perturbative stage is imposed. In particular,
a splitting of non-valence partons is enforced to allow for a smooth
transition to the valence partons inside the outer hadron. Although HERWIG
also allows for the simulation of a soft underlying event, we have not made
use of this possibility. Since the parton shower is supposed to describe
only the soft/collinear region, any initial-state emission outside this
region, i.e.\ where $\xi>z^2$, is suppressed.

As was the case for PYTHIA, hard matrix element corrections for SM vector
boson production have been implemented in HERWIG \cite{Corcella:1999gs},
where radiation in the region $\xi>z^2$ can now be allowed. It is then
distributed according to the matrix element describing the emission of an
additional parton. In the soft region $\xi<z^2$ already populated by the parton
shower, the emission of the hardest (largest $p_T$) parton generated so far is
reweighted in order to avoid double counting. Note that this need not be the
first emission, since angular ordering does not necessarily imply ordering in
transverse momentum. Note also that the normalization of the total cross
section is still accurate to LO only.

A NLO accuracy of the total cross section has been achieved in MC@NLO
\cite{Frixione:2002ik} through an implementation of NLO
cross sections, matched to the HERWIG parton shower. Instead of the LO
matrix element implemented in the standard version of HERWIG, two separate
samples of Born-like or standard ($S$) and hard emission ($H$) events are
generated, that can have weight $w_i^{(S,H)}=\pm 1$ and are statistically
distributed according to the positive-definite standard and hard
contributions to the NLO cross section $J_{S,H}$. These are made separately
finite by adding and subtracting the NLO part of the expanded Sudakov form
factor and are explicitly defined in Ref.\ \cite{Frixione:2002ik}. The
total cross section is then given by $\sigma_{\rm tot}=\sum_{i=1}^{N_{\rm
tot}} w_i^{(S,H)}(J_H+J_S)/N_{\rm tot}$.


\subsection{Joint transverse-momentum and threshold resummation for $Z'$-bosons}
\label{sec:2e}

The LO matrix element predictions for $Z'$-production are affected by fixed
order (F.O.) QCD corrections, that are logarithmically enhanced, when the
$Z'$-boson is produced close to the partonic threshold, i.e.\ $z=M^2/s\to1$,
or when its transverse momentum is small, i.e.\ $p_T\to0$. These
corrections must then be resummed (res) to all orders, which is most easily
achieved in Mellin ($N$) and impact parameter ($b$) space, and matched to
the F.O.\ prediction by subtracting from their sum the perturbatively
expanded (exp) resummed prediction, i.e.\
\bea
 \frac{\d^2\sigma}{\d M^2\,\d p_T^2} = \frac{\d^2\sigma^{({\rm
 F.O.})}}{\d M^2\,\d p_T^2} + \oint_{C} \frac{\d N}{2\pi
 i}\, \tau^{-N} \int_0^\infty \frac{b\, \d b}{2} J_0(b\, p_T)
 \left[\frac{\d^2\sigma^{{\rm (res)}}}{\d M^2\,\d p_T^2}(N, b)
 - \frac{{\rm d}^2\sigma^{{\rm (exp)}}}{\d M^2\,\d p_T^2}(N, b) \right].
\eea
In this way, a uniform precision is obtained, and the large theoretical
(renormalization and factorization scale) uncertainties are reduced.
The NLO cross section for Drell-Yan processes (with $\tau=M^2/S$)
\bea
 \frac{\d^2\sigma^{({\rm F.O.})}}{\d M^2\, \d p_T^2}&=&
 \sum_{a,b}\int_{\tau}^1 \!{\rm d}x_a \int_{\tau/x_a}^1 \!\!{\rm d}x_b\,
 f_{a/h_a}(x_a;\mu_F)\, f_{b/h_b}(x_b;\mu_F)
 \le \delta(p_T^2)\, \delta(1-z)\, \hat\sigma^{(0)}_{ab} +
 \frac{\alpha_s(\mu_R)}{\pi}\, \hat\sigma^{(1)}_{ab}(z) + {\cal
 O}(\alpha_{s}^{2})\re
\eea
is well-known \cite{Aurenche:1980tp}, and the necessary modifications for
implementing $Z'$-bosons in the ${\cal O}(\as^i)$ partonic cross sections
\bea
 \hat{\sigma}^{(i)}_{ab}&=&
 {1\over2s}
 \overline{|{\cal M}_{i}|^2}
 {\d t\over 8\pi s}
\eea
have already been discussed above.

Since the $p_T$- \cite{Bozzi:2006fw} and threshold-enhanced contributions
\cite{Bozzi:2007qr} are both due to soft-gluon emission in the initial
state, they may be resummed at the same time \cite{Li:1998is,Laenen:2000ij,%
Bozzi:2007te}. The logarithms are then organized by the function
\bea
 \chi(\bbar, \nbar)&=&\bbar + \frac{\nbar}{1+\eta\,\bbar/\nbar}
 ~~~{\rm with}~~~ \bbar \equiv b\, M\,e^{\gamma_E}/2
 ~~~{\rm and}~~~ \nbar \equiv N e^{\gamma_E},
 \label{eq:chi}
\eea
whose form with $\eta=1/4$ is constrained by the requirement that the
leading and next-to-leading logarithms in $\bbar$ and $\nbar$ are correctly
reproduced in the limits $\bar{b}\to \infty$ and $\bar{N}\to\infty$,
respectively. The choice of Eq.\ (\ref{eq:chi}) with $\eta = 1/4$ avoids the
introduction of sizeable subleading terms into perturbative expansions of
the resummed cross section at a given order in $\as$, which are not present
in fixed-order calculations \cite{Kulesza:2002rh}. The resummed cross
section
\bea
 \frac{\d^2\sigma^{({\rm res})}}{\d M^2\, \d p_T^2} (N, b)
 &=& \sum_{a,b,c}
 f_{a/h_a}(N+1; \mu_F)\, f_{b/h_b}(N+1; \mu_F) \,
 \hat{\sigma}_{c \bar{c}}^{(0)}
 \exp\le \mathcal{G}_c(N,b; \alpha_s, \mu_R)\re\nonumber\\
 &\times&
 \left[\delta_{ca} \delta_{{\bar c}b} + \sum_{n=1}^{\infty}\lr
 \frac{\alpha_{s}(\mu_R)}{\pi}\rr^{n}\, \mathcal{H}_{ab\to
 c\bar{c}}^{(n)}\Big(N; \mu_R, \mu_F\Big)\right]
\eea
can then be factorized into a regular part with
\bea
 \mathcal{H}_{ab\to c\bar{c}}^{(1)}\Big(N;\mu_R,\mu_F\Big) =
 \delta_{ca} \delta_{{\bar c}b}\, H_{c\bar{c}}^{(1)}(\mu_R) +
 \delta_{ca}\, C_{{\bar c}/b}^{(1)}(N) + \delta_{{\bar c}b}\,
 C_{c/a}^{(1)}(N) + \left( \delta_{ca} \gamma_{{\bar c}/b}^{(1)}(N) +
 \delta_{{\bar c}b} \gamma_{c/a}^{(1)}(N) \right)
 \ln\frac{M^2}{\mu_F^2},
\eea
where in the Drell-Yan resummation scheme
\bea
 H_{c\bar{c}}^{(1)}(\mu_R)~\equiv~0,~~
 C_{q/q}^{(1)}(N) = \frac{2}{3\, N\, (N+1)} + \frac{\pi^2-8}{3},
 &{\rm and}&
 ~~ C_{q/g}^{(1)}(N) = \frac{1}{2\, (N+1)\, (N+2)},
\eea
and a perturbatively calculable eikonal factor
\bea
 \mathcal{G}_c(N,b; \alpha_s, \mu_R) &=&
 g_c^{(1)}(\lambda)\,\ln\chi + g_c^{(2)}(\lambda; \mu_R),
\eea
which depends through the functions
\bea
 g_c^{(1)}(\lambda)&=& \frac{A_c^{(1)}}{\beta_0} \frac{2\, \lambda
 + \ln\big(1 - 2\, \lambda\big)}{\lambda} ~~{\rm and}\\
 g_c^{(2)}(\lambda; \mu_R) &=& \frac{A_c^{(1)}\,
 \beta_1}{\beta_0^3} \left[ \frac{1}{2} \ln^2 \big
 (1 - 2\, \lambda\big) + \frac{2\,
 \lambda + \ln\big(1 - 2\, \lambda \big)}{1 - 2\, \lambda} \right]\nn\\
 &+& \left[ \frac{A_c^{(1)}}{\beta_0} \ln\frac{M^2}{\mu_R^2} -
 \frac{A_c^{(2)}}{\beta_0^2} \right] \left[ \frac{2\, \lambda}{1 -
 2\, \lambda}+ \ln\big(1 - 2\, \lambda \big) \right] +
 \frac{B_c^{(1)}(N)}{\beta_0}\, \ln \big(1 - 2\, \lambda
 \big)
\eea
on the logarithm $\lambda = \beta_0/\pi\, \as(\mu_R)\ln\chi$. The anomalous
dimensions $\gamma_{c/a}^{(1)}(N)$ are the $N$-moments of the ${\cal O}
(\alpha_s)$ Altarelli-Parisi splitting functions. Up to next-to-leading
logarithmic order, the coefficients needed in $g^{(1,2)}_c$ are
\bea
 A_q^{(1)}~=~C_F,~~
 A_q^{(2)}~=~C_F\le C_A\lr{67\over36}-{\pi^2\over12}\rr
 -{5\over9}T_RN_F\re,&{\rm and}&
 B_q^{(1)}(N)~=~-{3\over2}C_F+ 2 \gamma_{q/q}^{(1)}(N).
\eea
The usual coefficients of the QCD $\beta$-function are
\bea
 \beta_0 = \frac{1}{12}(11\, C_A - 4\,T_R\, N_f) ~~{\rm and}~~
 \beta_1 = \frac{1}{24}(17\, C^2_A - 10\,
 T_R\, C_A\, N_f - 6\, C_F\, T_R\, N_f),
\eea
the number of effectively massless quark flavours is $N_f$, and
$C_F = 4/3$, $C_A = 3$, and $T_R=1/2$ are the usual QCD colour factors.
Re-expanding the resummed cross section leads to
\bea
 \frac{\d^2\sigma^{({\rm exp})}}{\d M^2\, \d p_T^2}(N,b)
 &=& \sum_{a,b,c} f_{a/h_a}(N+1; \mu_F)\, f_{b/h_b}(N+1;
 \mu_F)\, \hat{\sigma}_{c{\bar c}}^{(0)} \nonumber\\
 &\times&\le \delta_{ca} \delta_{{\bar c}b} +
 \sum_{n=1}^{\infty}\left(\frac{\as(\mu_R)}{\pi} \right)^n
 \lr\Sigma_{ab\to c{\bar c}}^{(n)}(N,b)+
 \mathcal{H}_{ab\to c\bar{c}}^{(n)}(N; \mu_R, \mu_F)\rr\re
\eea
with the coefficient
\bea
 \Sigma_{ab\to c{\bar c}}^{(1)}(N,b) &=& -2
 \le A^{(1)}_c \delta_{ca} \delta_{{\bar c}b} \ln^2\chi +
 \lr B^{(1)}_c \delta_{ca} \delta_{{\bar c}b} +
     \delta_{c       a} \gamma_{{\bar c}/b}^{(1)}(N) +
     \delta_{{\bar c}b} \gamma_{      c /a}^{(1)}(N)
                                          \rr \ln  \chi \re,
\eea
which matches precisely with the NLO cross section.

\section{Numerical comparison}
\label{sec:3}

In this section, we compare the three different theoretical approaches to
$Z'$-boson production at the LHC discussed above, i.e.\ the matching of
LO matrix elements with parton showers as implemented in PYTHIA, the
matching of NLO matrix elements with parton showers in our modified version
of MC@NLO, and the matching of NLO matrix elements with our improved
formalism of joint resummation at next-to-leading logarithmic (NLL) order.
For a comparison of joint resummation with transverse-momentum and
threshold resummation in Drell-Yan like processes we refer the reader to
Ref.\ \cite{Bozzi:2007te}.

We first fix our choice of parton densities and of the strong, electroweak,
and extra gauge-boson parameters and demonstrate that the three theoretical
predictions coincide at the LO partonic level for both the mass- and the
$p_T$-spectrum. Next, we show the impact of the dominant electroweak
corrections by running the fixed electromagnetic coupling from zero momentum
transfer to the mass of the $Z'$-boson. We then analyze separately the
different levels of improvement made possible in the three theoretical
approaches, before comparing the three ``best versions'' directly
with each other. Finally, we discuss the impact of the remaining theoretical
uncertainties, coming from variations of the renormalization and
factorization scales and the parton densities, on the theoretical
predictions.

\subsection{Choice of strong, electroweak, and extra gauge-boson parameters}

We will make predictions for $pp$ collisions at the LHC at a centre-of-mass
energy of $\sqrt{S}=14$ TeV, choosing the CTEQ6L (LO) and CTEQ6M (NLO
$\overline{\rm MS}$)  \cite{Pumplin:2002vw} sets as our default for the
parton densities in the protons for LO and NLO/NLL calculations,
respectively. The strong coupling constant $\alpha_s(\mu_R)$ is always
computed with two-loop accuracy, $\Lambda_{\overline{\rm
MS}}^{n_f=5}=226$ MeV, and identifying the renormalization scale $\mu_R$ (as
well as the factorization scale $\mu_F$) with the invariant mass of the
lepton pair $M$. As is customary, the theoretical uncertainty is estimated
by varying the scales by a factor of two around the central value.

For the electroweak mass, width, and coupling parameters, we use the values
of the 2002 Review of the Particle Data Group \cite{Hagiwara:2002fs}, i.e.\
$m_Z=91.188$ GeV, $\Gamma_Z=2.4952$ GeV, $\alpha=1/137.04$, and $\sin^2
\theta_W=0.23113$, which are (still) used as default in the $Z'$ analysis of
the ATLAS collaboration. The only value that has changed in 2006 is $\sin^2
\theta_W=0.23122$ \cite{Yao:2006px}, but the numerical impact of this change
remains visibly small. Using these parameters, a $Z'$-boson mass of 1 TeV,
running the fine-structure constant to $\alpha$(1 TeV)=1/124.43, and
including the NLO QCD correction factor $1+\alpha_s(\mu_R)/\pi$ for
$Z'$-decays into quarks, we compute the total width of the $Z'$-boson in the
$\chi$-model within PYTHIA (see Sec.\ \ref{sec:2.1}) to be $\Gamma=12.04$
GeV. Its branching ratio to electron-positron pairs, representing the
signal, is found to be 5.98\%. These values are then passed as parameters to
our modified MC@NLO and new resummation programs. 

\begin{figure}
 \centering
 \epsfig{file=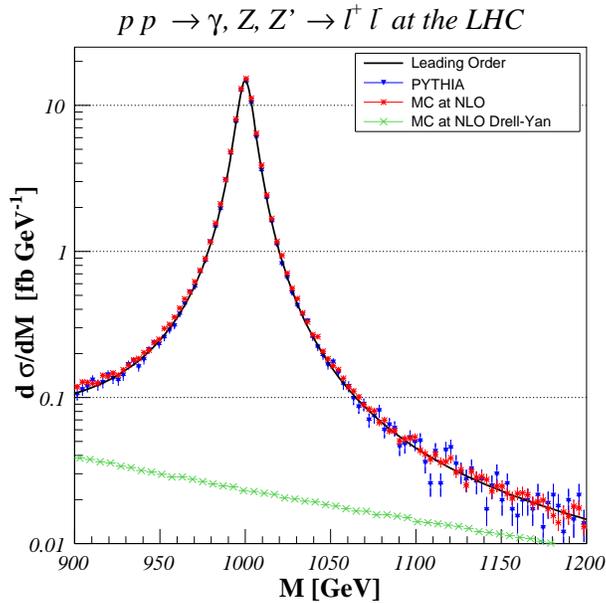,width=0.49\columnwidth}
 \caption{\label{fig:1}The LO mass-spectrum for
 $Z'$ production with fixed $\alpha$ in PYTHIA (triangles), MC@NLO (stars)
 and at parton level (full line and circle), compared to the SM Drell-Yan
 background in MC@NLO (crosses).}
\end{figure}

In order to set a common theoretical basis, we show in Fig.\ \ref{fig:1}
%
%
the invariant mass-spectrum of the electron-positron pair in LO QCD, i.e.\
without any parton
shower, matrix element correction, resummation, or hadronization effects.
The PYTHIA (triangles), MC@NLO (stars), and parton level (full line)
mass spectra, shown in the mass range of 900 to 1200 GeV around the
mass peak of the $Z'$-boson at 1 TeV, coincide perfectly. For comparison, we
also show the differential cross section for the SM Drell-Yan process in
MC@NLO (crosses), which coincides with the corresponding LO and PYTHIA
predictions and represents the dominant (irreducible) background to
the $Z'$-boson signal. Far from the resonance region, the two mass spectra
would, of course, coincide. The transverse momentum of the lepton pair
($p_T$) is exactly zero in PYTHIA and at parton level, while the forced
splitting of non-valence partons in MC@NLO induces a distribution that extends
to non-zero values of $p_T$ even when the parton shower is switched off.
Since this distribution is unphysical, we do not show it here. When
integrated over $p_T$, the MC@NLO total cross section coincides, however, with
PYTHIA's and the one at parton level.

The dominant electroweak corrections can be resummed by running the fixed
value of the fine-structure constant $\alpha=1/137.04$ in Fig.\ \ref{fig:1}
to the value $\alpha=1/124.43$ at the mass of the $Z'$-boson in the
$\overline{\rm MS}$-scheme. The ratio $\alpha^2(M)/\alpha^2(0)$ then 
induces an increase of about 22\% in the cross sections shown in Fig.\
\ref{fig:1}. In the following, we will always use a running value of
$\alpha$.

\subsection{Numerical results with PYTHIA}

\begin{figure}
 \centering
 \epsfig{file=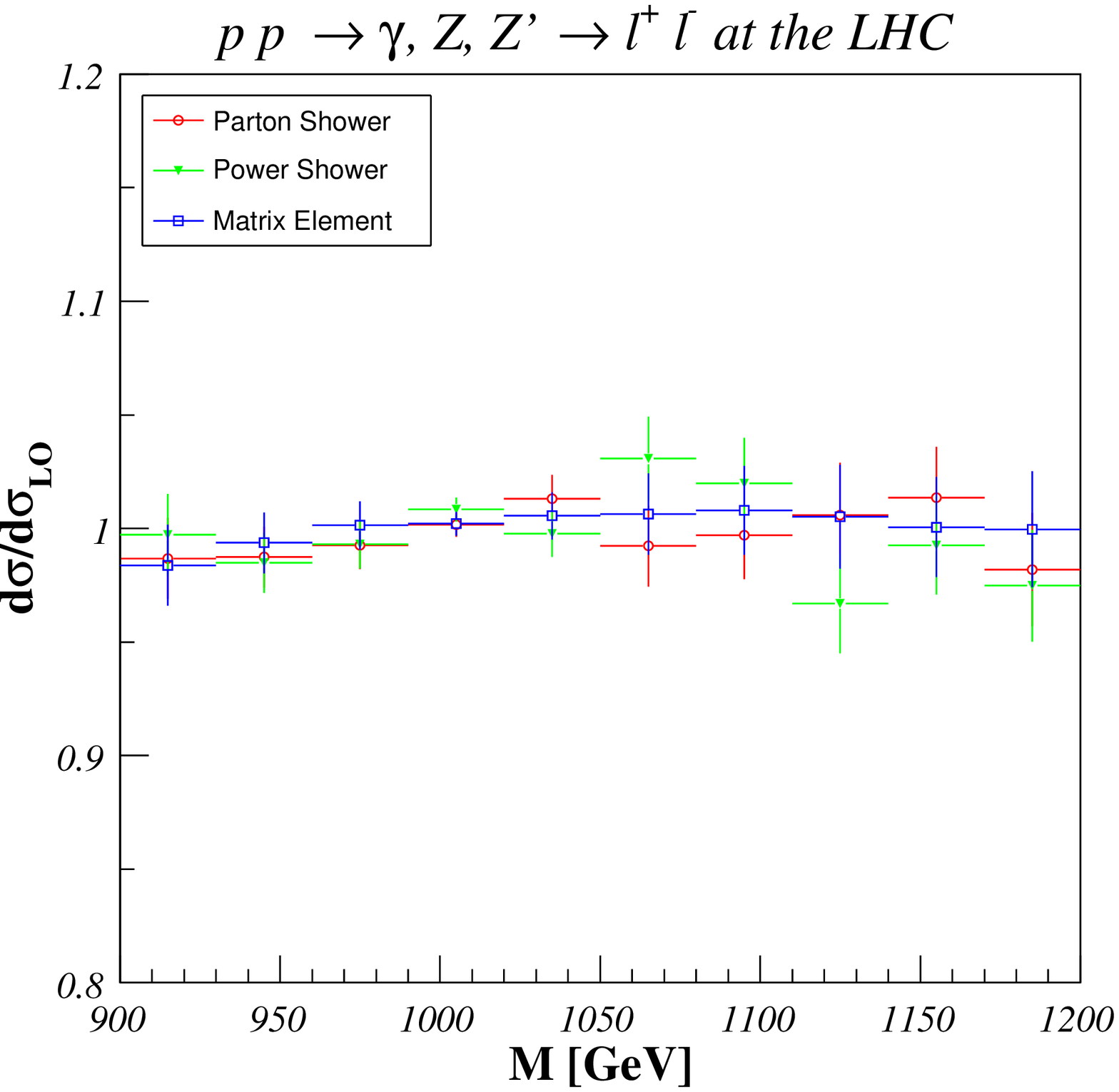,width=0.49\columnwidth}
 \epsfig{file=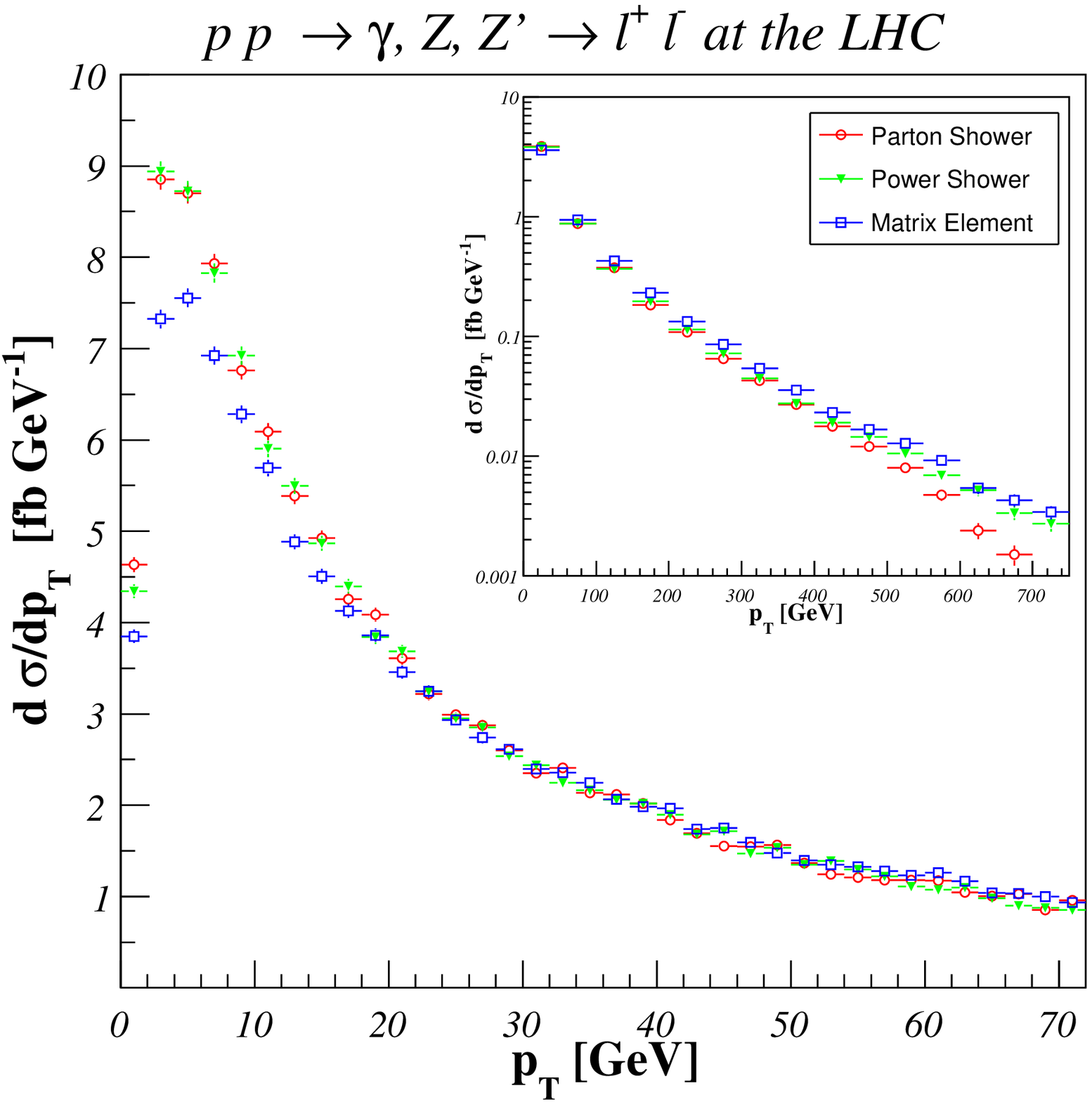,width=0.49\columnwidth}
 \caption{\label{fig:2}Mass (left) and transverse-momentum spectra (right)
 with PYTHIA with soft/collinear QCD parton showers (circles), QCD parton
 showers populating the full phase space (triangles), and after adding LO
 matrix element corrections (squares). The mass spectra have been normalized
 to the LO QCD prediction.}
\end{figure}

In Fig.\ \ref{fig:2}, we show the three different ways of improving on the
%
%
parton-level predictions that are implemented in PYTHIA, where one can
either add QCD parton showers in the soft and collinear regions (circles),
corresponding to the leading-logarithmic (LL) approximation, or in the full
phase space (triangles), which however overestimates the cross section, so
that it must be renormalized to the matrix element describing the emission
of an additional hard parton (squares). By definition, neither the total
cross section nor the mass spectrum are changed and remain accurate to LO
only, as can be seen from the fact that all histograms on the left-hand side
of Fig.\ \ref{fig:2} coincide, after normalization to the LO prediction and
within the statistical error bars, with unity. The right-hand side of Fig.\
\ref{fig:2} demonstrates that the zero transverse momentum of the $Z'$-boson
is smeared by the parton showers. It then peaks around 3 GeV and extends up
to 700 GeV (see insert) and even beyond, if the available phase space is
opened to the full hadronic centre-of-mass energy as proposed by the ``power
shower'' prescription. Note that strictly speaking the LL approximation is,
of course, no longer valid there. Normalizing the ``power shower'' to the
correct QCD matrix element describing one real emission brings the
integrated cross section back into agreement with the LO prediction, as the
increase of the cross section at large $p_T$ is compensated by a reduction
at small $p_T$.
These predictions may be improved further in PYTHIA by adding QED parton
showers, which are, however, formally suppressed by a factor of $\alpha/
\alpha_s$, so that their influence would not be visible in Fig.\
\ref{fig:2}. The emitted additional partons may furthermore hadronize, which
is modeled in PYTHIA with the Lund string model, but we have checked again
that the shapes of the distributions in Fig.\ \ref{fig:2} would not change
significantly.

\subsection{Numerical results with MC@NLO}

\begin{figure}
 \centering
 \epsfig{file=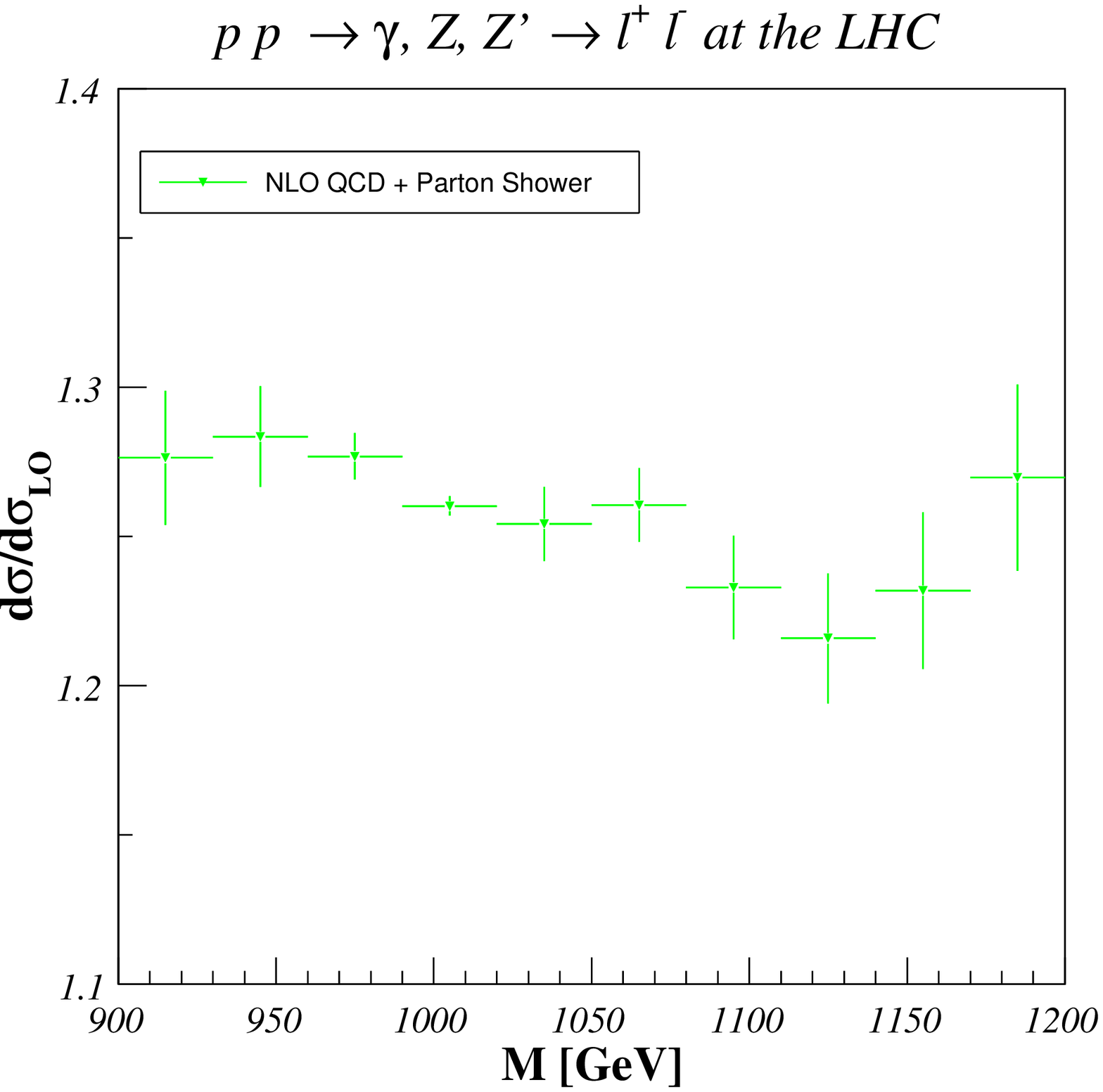,width=0.49\columnwidth}
 \epsfig{file=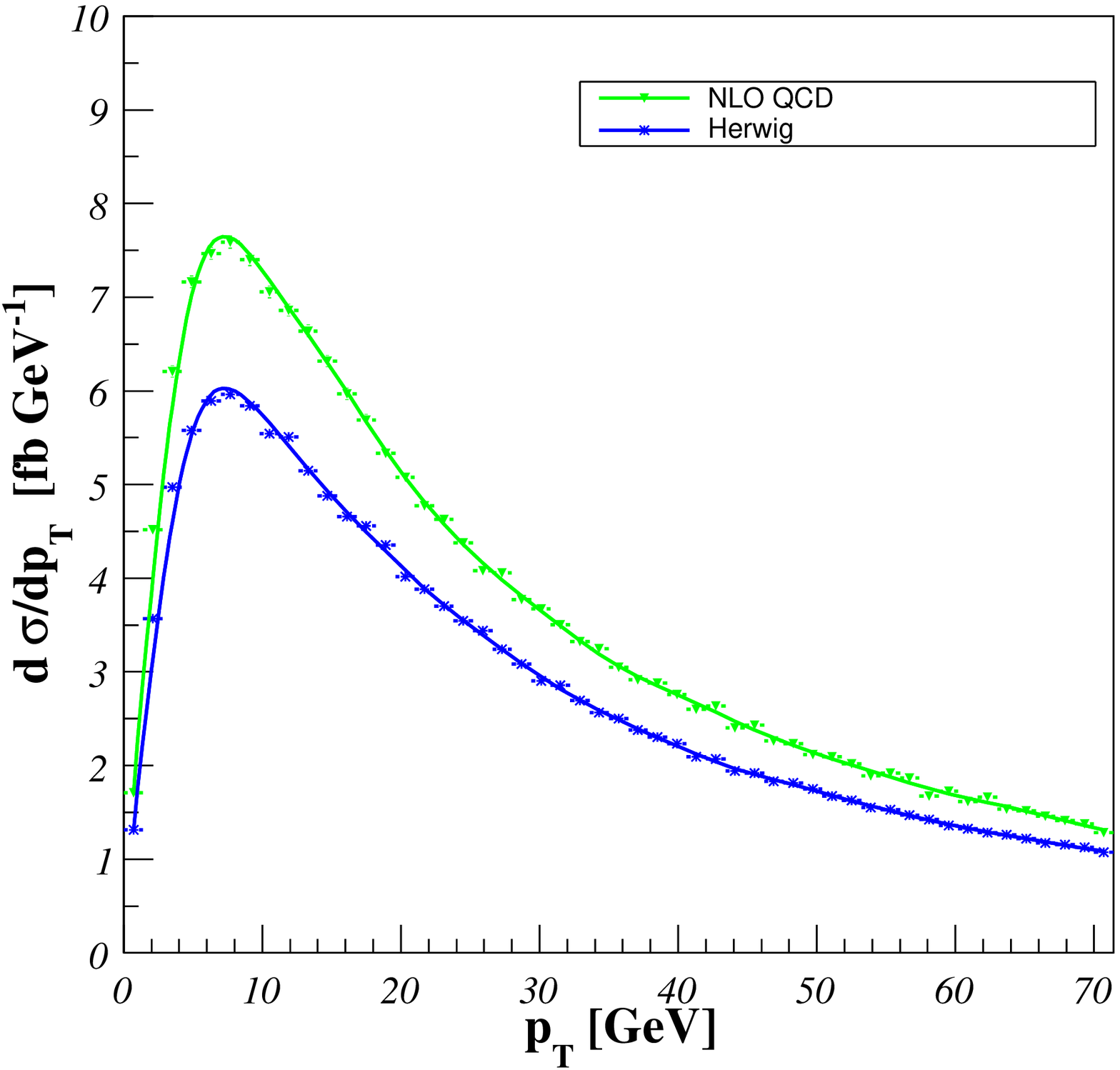,width=0.49\columnwidth}
 \caption{\label{fig:3}Mass (left) and transverse-momentum spectra (right)
 after matching the NLO QCD corrections
 to the HERWIG QCD parton shower (triangles). The mass spectra have been
 normalized to the LO QCD prediction.}
\end{figure}

With MC@NLO it is possible to correct not only the transverse-momentum
shape, but also the normalization of the total cross section. In our
implementation, this is now also possible for $Z'$ production at the LHC.
This is demonstrated in Fig.\ \ref{fig:3}, where we show on the left-hand
side that the NLO correction
%
%
factor 
\bea
 K&\equiv&{\d\sigma_{\rm NLO}\over\d\sigma_{\rm LO}}~\simeq~1.26
\eea
at the resonance ($M=1$ TeV) is quite significant and depends also
slightly on the invariant mass of the lepton pair. By definition, the
normalization remains again unchanged by the HERWIG parton shower, which
affects, however, strongly the $p_T$-spectrum (right). While the fixed-order
prediction diverges as $p_T\to0$, its logarithmic singularity is
effectively resummed by the parton shower and leads to a smooth turnover
with a maximum at around 8 GeV, i.e.\ at a value that is considerably larger
than in the case of PYTHIA. Note that the parton shower in MC@NLO
replaces part of the NLO contribution, so that switching it off (not shown)
does not lead to a fully correct NLO prediction.
As was already the case in PYTHIA,
we have checked that adding QED parton showers or hadronization, which is
modeled in HERWIG with the cluster model, does not lead to visible changes
in the distributions of Fig.\ \ref{fig:3}.

\subsection{Numerical results with joint resummation}

\begin{figure}
 \centering
 \epsfig{file=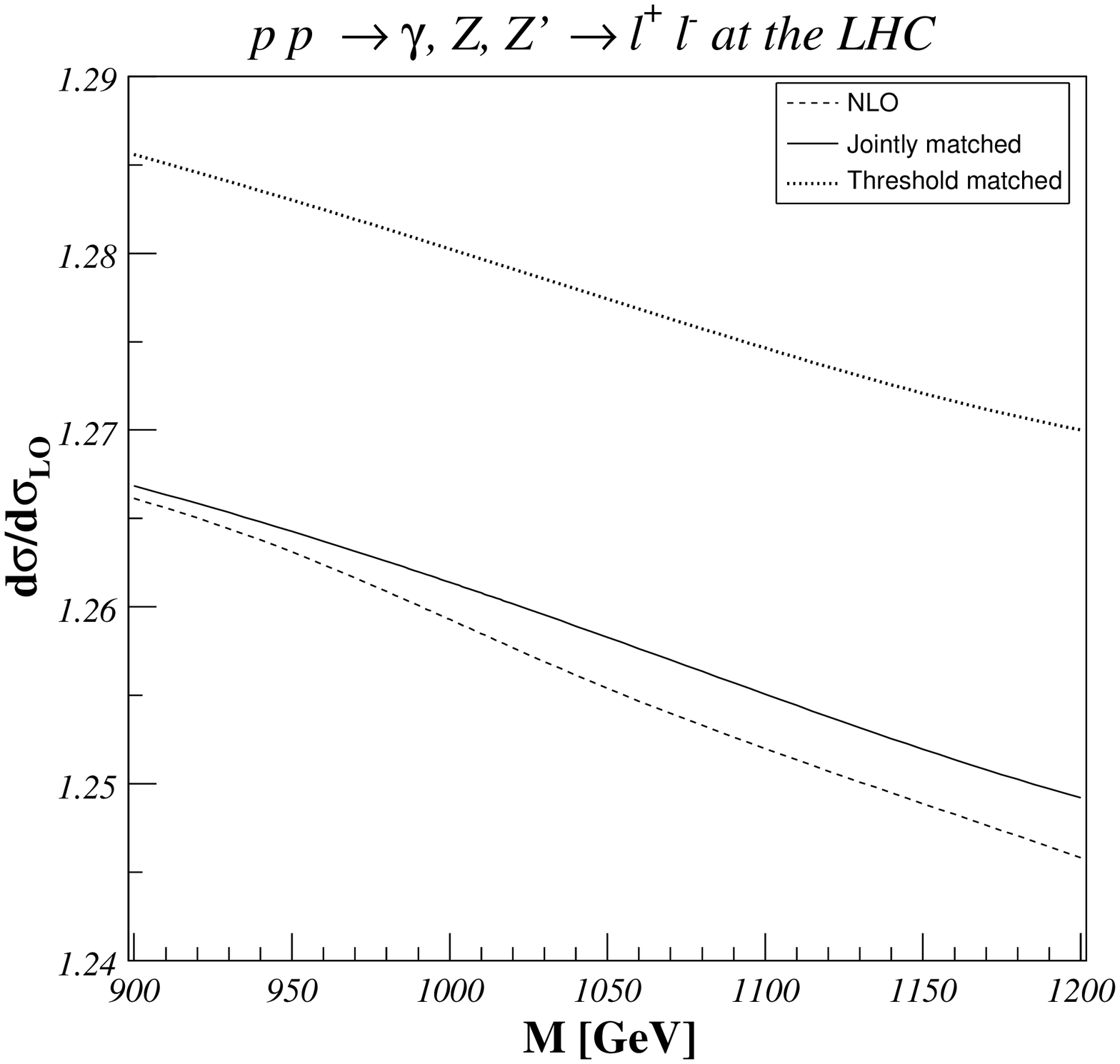,width=0.49\columnwidth}
 \epsfig{file=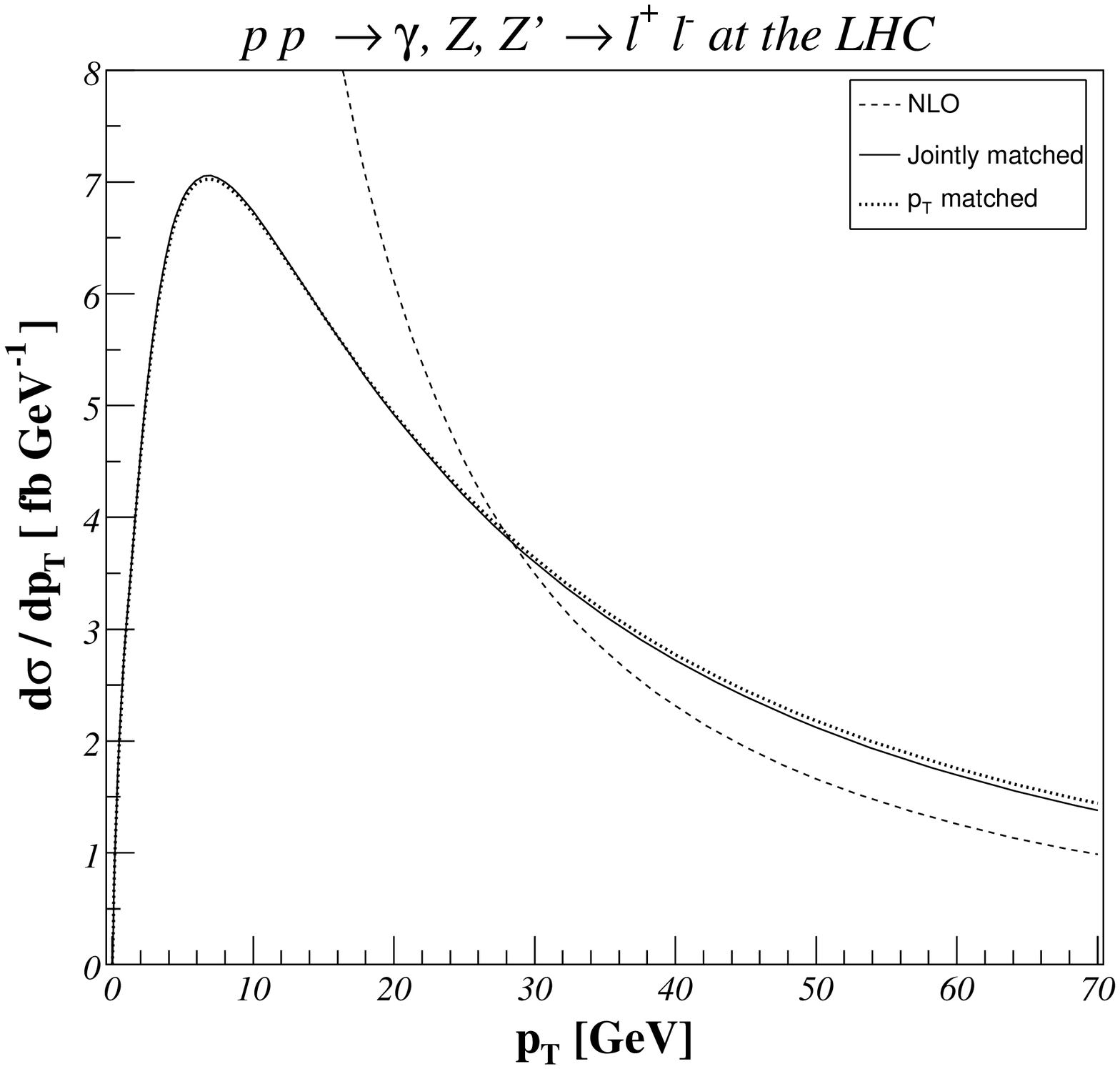,width=0.49\columnwidth}
 \caption{\label{fig:4}Mass (left) and transverse-momentum spectra (right)
 in NLO QCD (dashed) and after resumming threshold and $p_T$ logarithms
 (dotted) or both at the same time (full line). The resummed cross sections
 have been matched to those at NLO, and the mass spectra have been
 normalized to the LO QCD prediction.}
\end{figure}

Theoretically, the most precise predictions are obtained if logarithmically
enhanced soft and collinear parton emission is analytically resummed, both
close to threshold and close to $p_T\simeq0$, as described in Sec.\
\ref{sec:2e} above. As can be seen from Fig.\ \ref{fig:4} (left), the
%
%
NLO $K$-factor (dashed) is increased further by the resummed contributions,
in particular in pure threshold resummation (dotted), which is only
approximately described by joint resummation, as we are still relatively
far from the production threshold at $\sqrt{S}=14$ TeV. Only for larger
values of $M$ would the joint and threshold resummed predictions coincide.
As the additional increase in $K$ is only of the order of a few percent, the
prediction is nicely stabilized. We have checked that the $K$-factors of
1.26 at NLO and 1.28 at NNLO \cite{Hamberg:1990np} for Drell-Yan lepton
pairs with mass 1 TeV coincide precisely with our NLO and threshold resummed
results. Fig.\ \ref{fig:4} (right) demonstrates that the true
NLO $p_T$-distribution (dashed) diverges as $p_T\to0$, as it must, and does
indeed not exhibit the unphysical maximum of MC@NLO without parton showers
around 10 GeV. Resummation leads again to a smooth turnover with a maximum
at around 8 GeV. Note that the jointly resummed prediction (full) follows
the one with $p_T$-resummation (dotted) very closely, as is to be expected
from the definition of the variable $\chi$ and the form of the function
${\cal G}_c$ (see above).

\subsection{Comparison of numerical results and theoretical uncertainties}

\begin{figure}
 \centering
 \epsfig{file=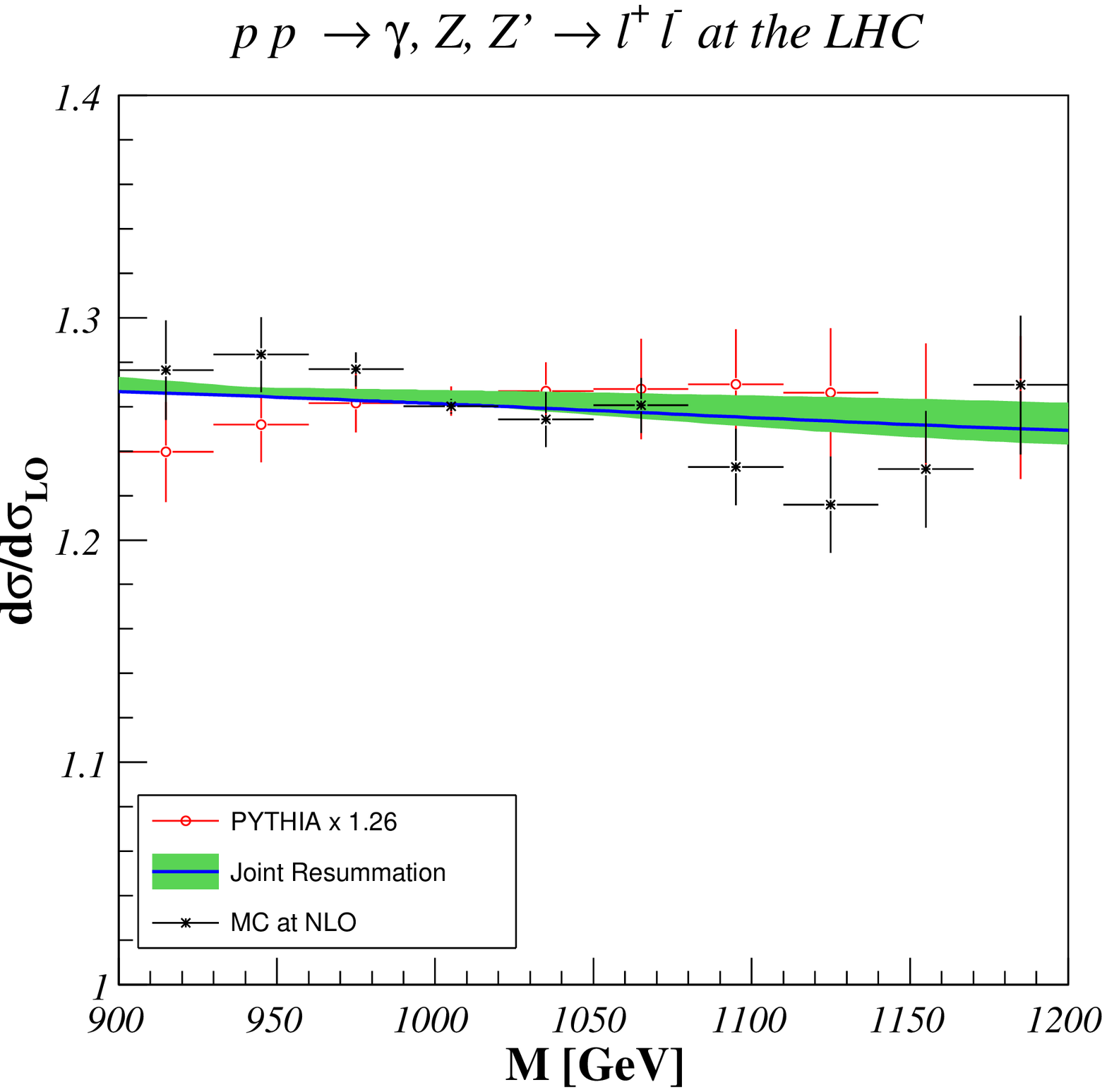,width=0.49\columnwidth}
 \epsfig{file=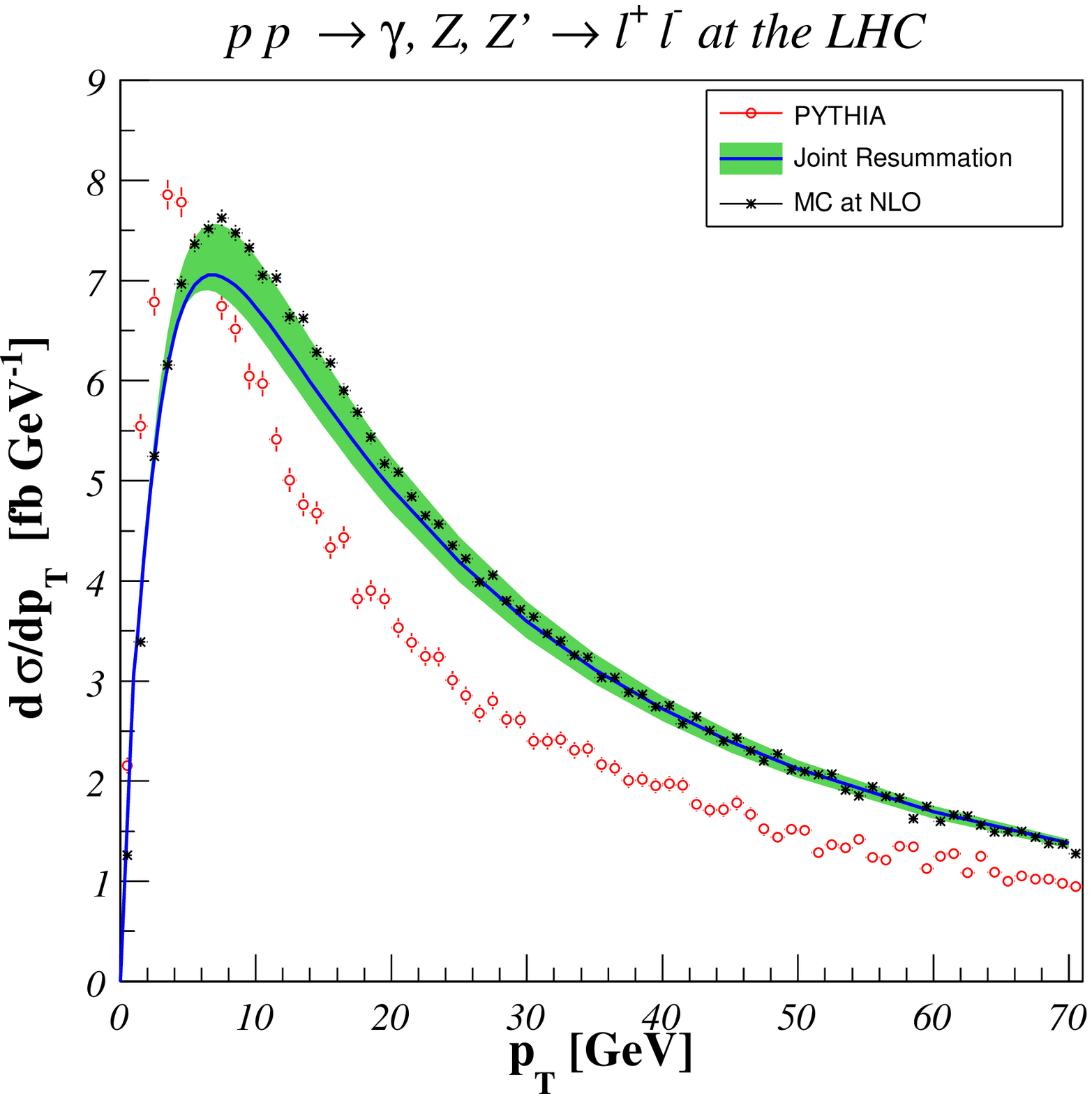,width=0.49\columnwidth}
 \caption{\label{fig:5}Mass (left) and transverse-momentum spectra (right)
 in PYTHIA with LO matrix elements matched to QCD parton showers (circles),
 in MC@NLO with NLO matrix elements matched to the HERWIG QCD parton shower
 (stars), and after matching the NLO QCD corrections to joint resummation
 (full line). The mass spectra have been normalized to the LO QCD
 prediction, and the renormalization and factorization scale uncertainties
 in the resummed predictions are indicated as shaded bands.}
\end{figure}

We now confront the ``best versions'' of the three different theoretical
approaches with each other, superimposing in Fig.\ \ref{fig:5} the LO
%
%
matrix element corrected predictions obtained with PYTHIA's ``power shower''
(circles), the NLO matrix element corrected predictions obtained with
MC@NLO's parton shower (stars), and the jointly resummed prediction matched
to the NLO matrix elements (full line). The correction factors for the mass
spectra (left), which have been normalized to the LO QCD prediction, show
only a very weak mass dependence. We have multiplied the PYTHIA mass
spectrum by hand with a global $K$-factor of 1.26. Otherwise, within the
statistical error bars the PYTHIA $K$-factor would just be unity, since the
normalization of the total cross section is changed
neither by the parton shower nor by the LO matrix element correction, that
serves to bring the ``power shower'' back into agreement with the LO QCD
prediction. The MC@NLO $K$-factor agrees almost perfectly with the one of
joint resummation, since we saw in Fig.\ \ref{fig:4} (left) that threshold
and joint resummation lead only to a very modest increase of the NLO
$K$-factor. The theoretical uncertainty induced in the resummed prediction
through the simultaneous variation of the renormalization and factorization
scale by a factor of two around the central scale $M$ is also considerably
smaller than the $K$-factor, indicating a nice stabilization of the
theoretical prediction. The $p_T$-spectra (right) are, for all three
improvements, no longer divergent. The PYTHIA prediction rises and falls
rather steeply around its maximum at 3 GeV, whereas the MC@NLO and resummed
predictions rise and fall more slowly around the peak at 8 GeV, which has
furthermore a slightly smaller cross section. The agreement between MC@NLO
and joint resummation is impressive, in particular for a scale choice of
$M/2$ (upper end of the shaded band) at low $p_T$ and $M$ at intermediate
$p_T$. We can therefore conclude that
our implementation of $Z'$ production in MC@NLO reaches almost the same
level of precision as our joint resummation calculation, but offers the
additional advantage of an easy implementation in the analysis chains of the
LHC experiments.

\begin{figure}
 \centering
  \epsfig{file=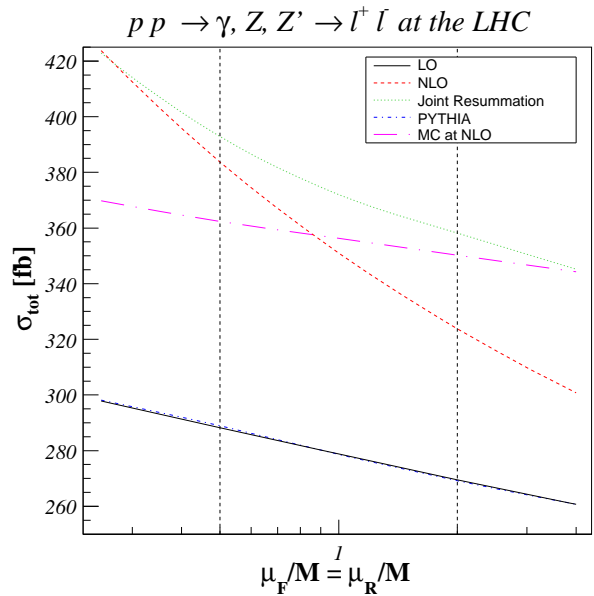,width=0.49\columnwidth}
 \caption{\label{fig:6}Dependence of the total $Z'$-boson production cross
 section at the LHC on the common factorization/renormalization scale
 $\mu_{F,R}$ in LO QCD (full), NLO QCD (dashed), and after matching the NLO
 QCD corrections to joint resummation (dotted), LO matrix elements to the
 PYTHIA parton shower (dot-dashed), and NLO matrix elements to the HERWIG
 parton shower (long dot-dashed).}
\end{figure}

The scale uncertainty of the total cross section (integrated over all
transverse momenta and over the invariant mass in the range from 900 to 1200
GeV) is shown in Fig.\ \ref{fig:6}. The LO QCD prediction (full) agrees with
%
%
the PYTHIA prediction (dot-dashed) at the same order, as the total cross
section and its scale dependence is not modified by the parton shower. Since
$\alpha_s(\mu_R)$ does not enter the calculation at this order, the full
scale dependence is in fact due to the factorization scale. The LO scale
dependence does, however, not give a reliable estimate of the theoretical
error, since the NLO cross section (dashed) is considerably larger. At NLO,
the factorization scale dependence is reduced as expected, but $\alpha_s
(\mu_R)$ makes its appearance, so that an additional renormalization scale
dependence is introduced. The total NLO scale dependence is reduced to
9\% (vertical lines), once the leading and next-to-leading logarithms
(NLL) are resummed (dotted). The MC@NLO prediction (long dot-dashed) agrees
with the one at NLO (dashed) for the central scale, but has a weaker
scale dependence due to the resummation of leading logarithms in the parton
shower \cite{Frixione:2004us}. Since both the type ($p_T$ vs.\ joint) and
order (LL vs.\ NLL) of the resummed logarithms differ between MC@NLO and
joint resummation, the two scale variations need not (and do not) coincide.

\begin{figure}
 \centering
 \epsfig{file=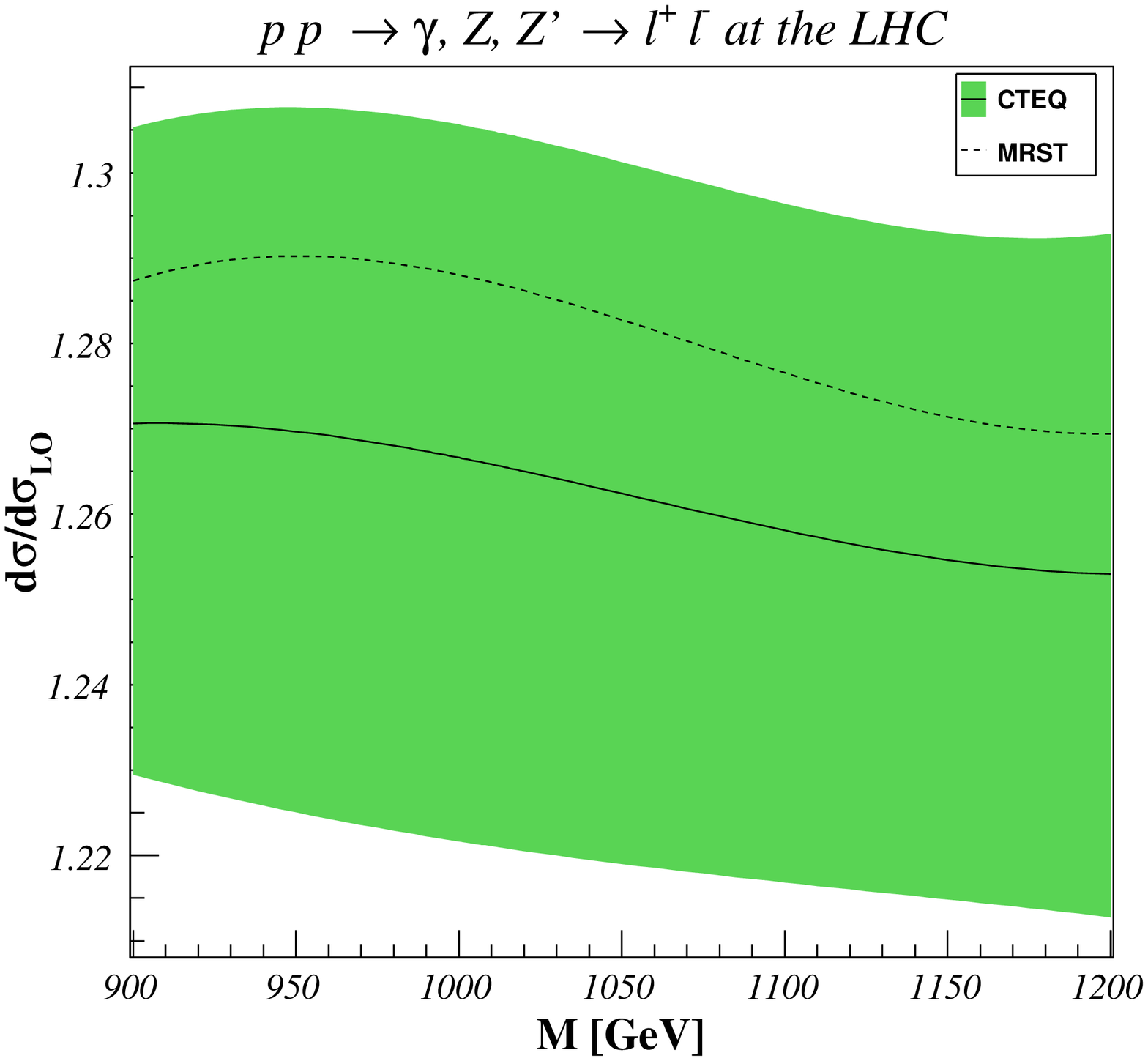,width=0.49\columnwidth}
 \epsfig{file=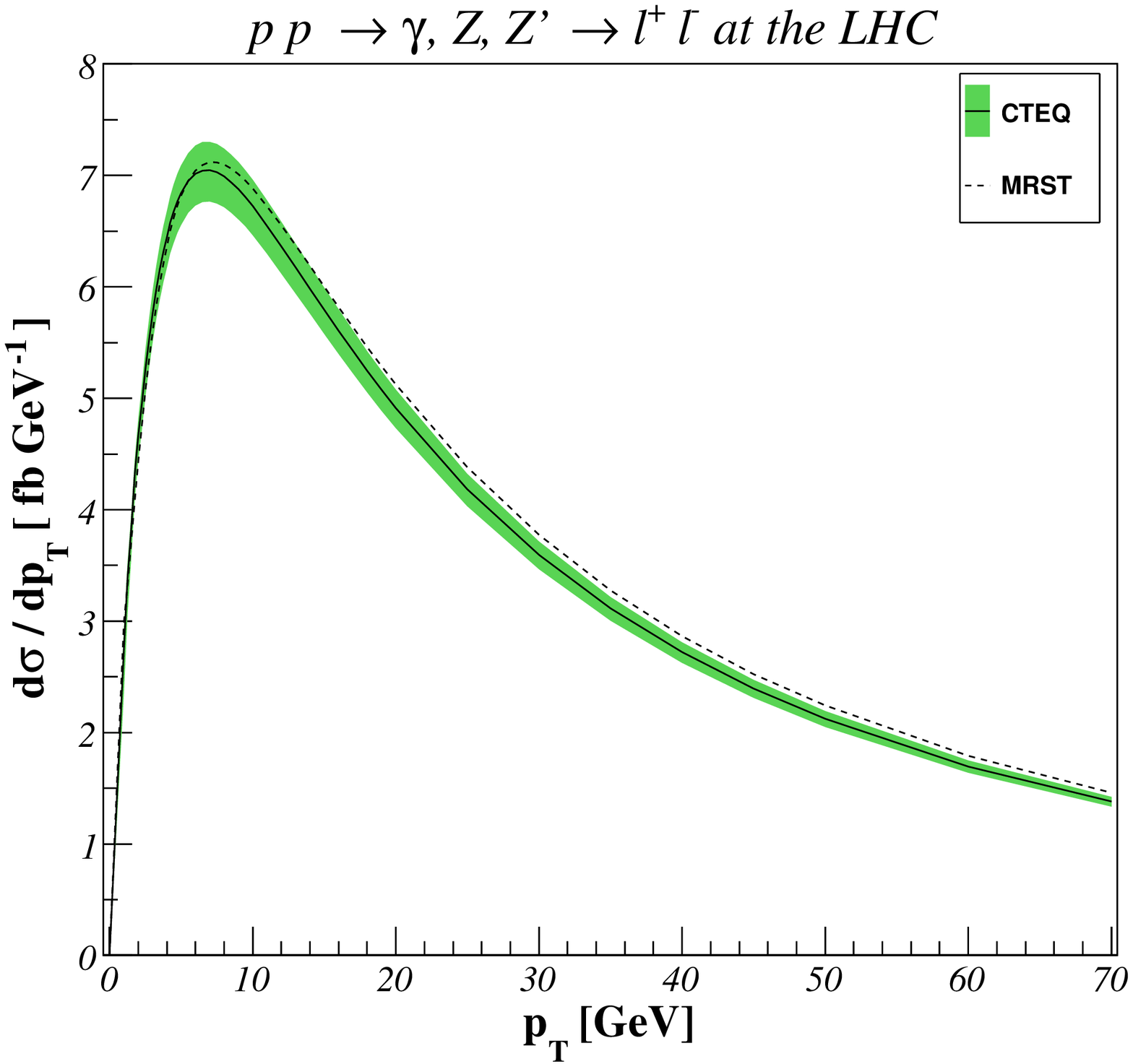,width=0.49\columnwidth}
 \caption{\label{fig:7}Mass (left) and transverse-momentum spectra (right)
 after matching the NLO QCD corrections to joint resummation with CTEQ6M
 (full) and MRST 2004 NLO \cite{Martin:2004ir} (dashed) parton densities.
 The mass spectra have been normalized to the LO QCD prediction using CTEQ6L
 and MRST 2001 LO \cite{Martin:2002dr} parton densities, respectively.
 The shaded bands indicate the maximal possible deviations allowed by the
 up and down variations along the 20 independent directions that span the
 90\% confidence level of the data sets entering the CTEQ6 global fit
 \cite{Tung:2006tb}.}
\end{figure}

We estimate in Fig.\ \ref{fig:7} the theoretical uncertainty
%
%
coming from different parameterizations of parton densities. Since the
invariant mass $M$ of the lepton pair is correlated with the momentum
fractions $x_{a,b}$ of the partons in the external protons, the normalized
mass spectra (left) are indicative of the different shapes of the quark and
gluon densities in the CTEQ6M (full) and MRST 2004 NLO \cite{Martin:2004ir}
(dashed) parameterizations. The latter also influence the
transverse-momentum spectra, which are slightly harder for MRST 2004 NLO
than for CTEQ6M. The shaded bands show the uncertainty induced by variations
along the 20 independent directions that span the 90\% confidence level of
the data sets entering the CTEQ6 global fit \cite{Tung:2006tb}. It
remains modest, i.e.\ about 8\%, and is thus slightly smaller than the scale
uncertainty of 9\%.

\begin{figure}
 \centering
 \epsfig{file=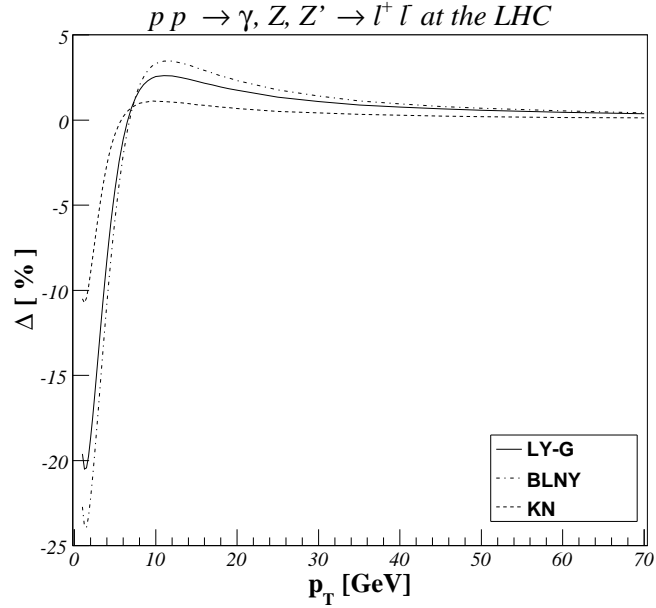,width=0.49\columnwidth}
 \caption{\label{fig:8}Variation (in percent) of the transverse-momentum
 spectrum after matching the NLO QCD corrections to joint resummation with
 CTEQ6M parton densities for three three different choices of a
 non-perturbative form factor.}
\end{figure}

The uncertainty at low transverse momenta coming from non-perturbative
effects in the PDFs is usually parameterized with a Gaussian form factor
describing the intrinsic transverse momentum of partons in the proton. We
show in Fig.\ \ref{fig:8}
%
%
the effects coming from three different choices of the form factor, i.e.\
those of Ladinsky-Yuan (LY-G) \cite{Ladinsky:1993zn},
Brock-Landry-Nadolsky-Yuan (BLNY) \cite{Landry:2002ix}, and
Konychev-Nadolsky (KN) \cite{Konychev:2005iy}, on the quantity
\bea
 \Delta &=& \frac{d\sigma^{\rm (res.+NP)}(\mu_R=\mu_F=M)-d\sigma^{(\rm
res.)}(\mu_R=\mu_F=M)}{d\sigma^{(\rm res.)} (\mu_R=\mu_F=M)}.
\eea
It is obvious that these
non-perturbative contributions are under good control, as their effect
is always less than 5\% for $p_T >$ 5 GeV and thus considerably smaller
than the scale and PDF uncertainties.

\section{Conclusions}
\label{sec:4}

In summary, we have improved the theoretical predictions for the production
of extra neutral gauge bosons at hadron colliders by implementing the $Z'$
bosons in the MC@NLO generator and by computing their differential and total
cross sections in joint $p_T$ and threshold resummation. The two improved
predictions were found to be in excellent agreement with each other for mass
spectra, $p_T$ spectra, and total cross sections, while the PYTHIA parton
and ``power'' shower predictions usually employed for experimental analyses
show significant shortcomings both in
normalization and shape. The theoretical uncertainties from scale and parton
density variations and non-perturbative effects were found to be 9\%, 8\%,
and less than 5\%, respectively, and thus under
good control. The implementation of our improved predictions in terms of the
new MC@NLO generator or resummed $K$ factors in the analysis chains of the
Tevatron and LHC experiments should be straightforward and lead to more
precise determinations or limits of the $Z'$ boson masses and/or couplings.
While we have shown numerical results for $Z'$ bosons associated with the
$U(1)_\chi$ gauge symmetry, our calculations are completely general and
easily applicable to different grand unification, extra-dimensional, or
other models. Our modified MC@NLO program has been endorsed by the original
authors and can be obtained, like the joint resummation program, at
{\tt http://lpsc.in2p3.fr/klasen/software}.

\acknowledgments
This work has been supported by the CNRS/IN2P3 with a pilot grant for the
LHC-Theory-France initiative and by two Ph.D.\ fellowships of the French
ministry for education and research. We thank B.\ Cl\'ement for finalizing
several figures, D.\ de Florian for allowing us to use his parton density
routines, and S.\ Frixione for useful comments on the manuscript.


\end{document}